\documentclass[conference]{IEEEtran}
\usepackage[numbers,sort&compress]{natbib}
\usepackage{booktabs}
\usepackage{longtable}
\usepackage{multirow}
\usepackage{graphicx}
\usepackage{amsmath}
\usepackage{amssymb}
\usepackage{xurl}
\usepackage{hyperref}
\usepackage{seqsplit}
\usepackage{multirow}
\graphicspath{{figures/}}
\hypersetup{hidelinks}

\newcommand{\hashvalue}[1]{%
  \texttt{\seqsplit{#1}}%
}

\newcommand{\longid}[1]{%
  \texttt{\seqsplit{#1}}%
}

\newcommand{\safeincludegraphics}[2][]{%
  \IfFileExists{#2}{%
    \includegraphics[#1]{#2}%
  }{%
    \fbox{%
      \parbox[c][0.22\textheight][c]{0.94\linewidth}{%
        \centering
        Missing figure file:\\[4pt]
        \texttt{\detokenize{#2}}%
      }%
    }%
  }%
}

\title{Evidence-Grounded LLM-Assisted Android Mobile Forensic Reporting}

\author{
\IEEEauthorblockN{
Madhab Chandra Das,
Sajid Ahmed,
Tasnia Tabassum,
Qingzhong Liu
}
\IEEEauthorblockA{
Department of Computer Science\\
Sam Houston State University\\
Huntsville, TX, USA\\
\{mxd147, sxa202, txt060, qxl005\}@shsu.edu
}
}

\begin{document}

\def\floatpagefraction{1}
\def\textpagefraction{.001}

\maketitle

\begin{abstract}
Large language models can assist forensic report drafting, but fluent output does not ensure evidentiary reliability. This study presents an evidence-grounded framework for evaluating LLM-assisted Android forensic reports at the atomic-claim level. Gemini, Llama, and Qwen each generated reports for two controlled cases: a public Android 13 dataset and a physical LG device with a zero-reportable-fact boundary. The six accepted reports were decomposed into 612 claims and evaluated against frozen, human-reviewed authorities through content labeling, citation validation, fact-recall analysis, independent rechecking, and dispute resolution. The final corpus contained 424 supported, 163 unsupported, four contradicted, four overstated or incomplete, and 17 excluded claims. In the public case, all three models recovered all 27 authorized report-level facts. Citation validity nevertheless differed substantially: Gemini produced 168 valid citations from 189 required citations, Qwen produced 145 from 189, and Llama produced none from 135. In the physical-device case, the models generated 65 claims despite the zero-reportable-fact boundary, including 38 unsupported and one contradicted claim. These results show that evidence restriction and complete fact recall do not guarantee claim support or valid evidence attribution. The framework provides an auditable method for detecting unsupported content, preserving claim-to-evidence lineage, and retaining investigator authority over final forensic reports.
\end{abstract}

\begin{IEEEkeywords}
Android forensics, digital evidence, forensic reporting, large language models, claim verification, reliability
\end{IEEEkeywords}

\section{Introduction}
\label{sec:introduction}

Mobile devices retain communication records, application data, files, timestamps, and other artifacts that may support a digital investigation. A forensically sound mobile examination must preserve evidence integrity throughout acquisition, examination, analysis, and reporting \citep{Ayers2014,Moreb2024}. Android investigations are challenging because applications use diverse storage structures and may change across operating-system and application versions. These conditions complicate artifact identification and interpretation, particularly when manual analysis is required \citep{Lin2018,Ryu2018}. The reporting stage introduces an additional reliability concern because every factual statement must remain consistent with the examined evidence and its documented meaning.

Artificial intelligence can assist with the analysis and presentation of digital evidence. However, its use in mobile forensics requires transparency, interpretability, robustness, traceability, and human oversight \citep{Vasilaras2024,Shingade2025,Alam2026}. These requirements are especially important when an AI-generated statement may influence an investigative conclusion. Large language models can generate readable summaries and structured report text, but linguistic fluency does not establish evidentiary support. \citet{michelet2024chatgpt} found that LLMs could assist forensic report writing only when their outputs received thorough proofreading and correction. ForensicLLM further showed that domain adaptation can improve correctness and source attribution in digital forensic question answering \citep{Sharma2025}. Nevertheless, generated findings and evidence references must still be verified before they are accepted in a forensic report.

Recent studies have examined complementary parts of this reliability problem. \citet{Studiawan2025} proposed a standardized methodology and ground-truth dataset for evaluating LLM-based forensic timeline analysis. \citet{Khalid2024} presented an explainable AI framework for digital forensic investigations and retained human validation within the investigative workflow. \citet{khatiwala2025evaluating} proposed a structured approach for linking AI-identified forensic artifacts to traceable identifiers and forensic cross-references. These studies address report assistance, source attribution, timeline evaluation, explainability, and evidence traceability. However, they do not jointly provide an atomic claim-level evaluation of complete Android forensic reports generated by multiple pretrained LLMs. They also do not compare reports from a public Android dataset and a physical Android device under the same generation, evidence-control, and scoring procedures.

This study addresses this gap through an evidence-grounded framework for evaluating LLM-assisted Android forensic reporting. The objective is to determine whether factual claims generated by Gemini, Llama, and Qwen remain within verified forensic evidence boundaries. The experiment uses two controlled cases with different reporting conditions. The \texttt{A13\_PUBLIC\_001} case is based on a public Android 13 dataset and contains a bounded set of authorized report-level facts. The \texttt{LG\_PHONE\_001} case is based on a physical-device acquisition and applies a controlled zero-reportable-fact boundary. Within each case, the three models received the same evidence package and reporting requirements.

Each model generated one accepted report for each case, producing six reports. The reports were decomposed into 612 atomic claims and evaluated against frozen, human-reviewed evidence authorities. The evaluation separately assessed claim support, contradiction, overstatement, exclusion status, citation validity, and authorized-fact recall. All claim labels underwent an independent recheck, and disagreements were resolved before the final scoring dataset was accepted. This design does not assume that evidence grounding eliminates hallucination. Instead, it identifies statements that remain within the accepted evidence boundary and statements that require correction or exclusion.

This study makes three contributions. First, it provides a controlled comparative evaluation of three pretrained LLMs across a public Android case and a physical-device case with distinct evidence boundaries. Second, it introduces an atomic claim-level evaluation method that separates evidentiary support, contradiction, overstatement, citation validity, and fact recall. Third, it preserves the scoring process through evidence lineage, independent rechecking, dispute resolution, and final analyst acceptance. Together, these contributions provide an auditable method for assessing LLM-assisted forensic reporting while retaining the investigator's authority over the final report.

The study answers the following research questions:

\begin{description}
    \item[\textbf{RQ1.}] What claim-level evidence-grounding and observed hallucination rates occur in the forensic reports generated by Gemini, Llama, and Qwen for the \texttt{A13\_PUBLIC\_001} and \texttt{LG\_PHONE\_001} cases?

    \item[\textbf{RQ2.}] How do the three models differ in their supported, unsupported, contradicted, and overstated or incomplete claims under the same evidence and reporting controls?

    \item[\textbf{RQ3.}] To what extent do the generated reports recover the authorized forensic facts and provide valid references to their supporting evidence?

    \item[\textbf{RQ4.}] How do claim-level reliability, fact recall, and citation validity differ between the public Android case and the physical-device case?
\end{description}

\section{Related Work}
\label{sec:related_work}

Prior research relevant to this study can be grouped into three areas: Android artifact examination, AI-assisted forensic analysis, and LLM reliability in forensic tasks.

Android forensic research has primarily focused on evidence acquisition, artifact discovery, and application-data interpretation. Established mobile forensic guidance emphasizes evidence preservation, controlled examination, and documentation throughout the investigative process \citep{Ayers2014}. \citet{Moreb2024} extended this process-oriented view through a mobile forensic investigation framework that connects acquisition, examination, and evidence management. At the artifact level, Fordroid automated the identification of sensitive information and database structures in Android applications \citep{Lin2018}. Scenario-based analysis has also shown that third-party applications leave useful traces, although frequent operating-system and application updates make artifact locations and meanings difficult to maintain \citep{Ryu2018}. More recent tools continue to address this problem. Argus monitors file-system changes caused by mobile application activity and supports the identification of relevant application files \citep{Boztas2025}. Controlled Android browser experiments have also shown that artifact recoverability and interpretation depend on the application, usage mode, storage source, and acquisition method \citep{Giannakopoulos2026}. These studies improve evidence discovery and interpretation, but they do not evaluate the factual reliability of complete reports generated from the recovered evidence.

A second research area examines the use of AI in mobile and digital forensics. \citet{Vasilaras2024} reviewed AI use in mobile forensics and identified interpretability, transparency, robustness, governance, and standardized procedures as important requirements. Other studies have similarly described AI as a means of supporting large-scale forensic analysis while retaining human control over investigative decisions \citep{Dwivedi2024}. Explainable AI frameworks have formalized this requirement. \citet{Khalid2024} proposed an XAI-based digital forensic workflow that includes human validation, while \citet{Shingade2025} examined trust, transparency, accountability, and forensic applicability in AI-supported incident response. \citet{Alam2026} further argued that explanations should be integrated throughout the cyber forensic lifecycle to support informed decisions and stakeholder trust. This work establishes the need for trustworthy and reviewable AI, but it mainly addresses analytical models and explanation mechanisms rather than the verification of factual statements in LLM-generated forensic reports.

LLM-specific research has demonstrated both practical value and evidentiary risk. \citet{michelet2024chatgpt} evaluated LLM-assisted forensic report writing and concluded that generated reports still required thorough proofreading and correction. \citet{Sharma2025} introduced ForensicLLM, a domain-specific local model for digital forensic question answering and source attribution. Its results showed that domain adaptation can improve correctness, relevance, and attribution. However, hallucination and source reliability remained important concerns. LLMs have also been applied to narrower forensic tasks. \citet{Studiawan2025} proposed a standardized methodology for evaluating LLM-generated forensic timelines against known ground truth. \citet{Pawlaszczyk2026} evaluated LLM-based SQL generation for heterogeneous mobile forensic databases, while \citet{Pawlaszczyk2025} used LLM agents to support controlled mobile forensic dataset creation. These studies evaluate specific outputs such as answers, timelines, SQL queries, or synthetic activity sequences. They do not evaluate complete forensic reports as collections of individually verifiable factual claims.

Reliability and traceability have recently received more direct attention. \citet{khatiwala2025evaluating} proposed a framework that links LLM-assisted artifact analysis to deterministic identifiers and forensic cross-references. This approach supports evidentiary consistency and auditability. However, artifact-level traceability alone does not establish whether every statement in a generated report is supported, contradicted, overstated, correctly cited, or omitted. A report can recover relevant facts while still containing unsupported interpretations or invalid evidence references.

The literature therefore provides important methods for Android artifact examination, AI-assisted analysis, explainability, source attribution, and task-specific LLM evaluation. The remaining problem is the controlled evaluation of complete LLM-generated Android forensic reports at the atomic claim level. This study addresses that problem by evaluating Gemini, Llama, and Qwen under the same case-specific evidence boundaries. It compares one public Android case and one physical-device case. Each generated report is decomposed into atomic claims and checked against a frozen, human-reviewed evidence authority. Claim support, contradiction, overstatement, citation validity, and authorized-fact recall are evaluated separately. Independent rechecking and dispute resolution are then used to produce the final scoring dataset. This design extends prior work from artifact discovery and task-level evaluation to an auditable assessment of the factual reliability of complete forensic reports.

\section{Dataset Construction}
\label{subsec:dataset_construction}

The study constructed two case-specific Android forensic corpora. The first corpus was derived from a public Android 13 dataset, and the second was derived from a physical LG device acquisition. For each case, evidence preservation, extraction, normalization, human review, experiment supply, and evaluation authority were completed as separate controlled stages. This separation preserved evidence integrity and documented each transformation applied during forensic processing \cite{Ayers2014}.

Neither corpus was used to train or fine-tune a language model. Each corpus served two distinct purposes: it provided a frozen evidence package for report generation and an independent evidence authority for subsequent claim evaluation. The two corpora were not merged because they differed in source, record structure, reportability policy, and evaluation role.

\subsubsection{Public Android Case}
\label{subsubsec:a13_construction}

The public corpus was constructed from the Digital Corpora Android 13 reference dataset contributed by Joshua Hickman \cite{DigitalCorporaAndroid13}. The extracted archive, \texttt{android\_13\_data.tar.gz}, was assigned the case identifier \texttt{A13\_PUBLIC\_001}. The preserved archive had the following SHA256 value:

\begin{center}
\hashvalue{AEB95FE9ED2DABEFEF786C3DC620B0CD92255476C7A9BE01049CB05D995A39F9}
\end{center}

The archive was processed with ALEAPP. All 388 configured parsers completed successfully. The processing stage produced 51,849 output files, including 266 TSV files and 221 database files.

The complete output inventory was reviewed before artifact selection. The review verified 361,058 logical TSV records and identified 33 source representations that required preservation during normalization. It also recorded 17 traceback markers across 13 artifact groups, two mixed-table anomalies, one verified duplicate export, and other conditional sources. Outputs with unresolved source or handling conditions were not admitted to the trusted corpus.

The accepted selection contained 26 artifacts. Twenty-four artifacts were classified as Tier 1A because they had verified TSV exports. Two artifacts were classified as Tier 1B because they required source-based handling. A further 22 entries remained excluded or conditional. These controls prevented incomplete, duplicate, mixed-table, and unresolved outputs from entering the normalized corpus.

The 26 accepted artifacts were normalized into 26 UTF-8 JSONL files. The normalized corpus contained 2,192 records with 2,192 unique record identifiers and preserved 199 multiline text values. Validation identified no record failures and no duplicate record identifiers.

Human review was completed in 26 controlled batches. Each batch contained 25 field decisions and two timestamp decisions. This process produced 650 resolved field decisions and 52 resolved timestamp decisions, for a total of 702 review decisions. The review distinguished values that could support forensic facts from values limited to traceability, technical metadata, structural content, exclusion, or uncertainty.

The accepted ground-truth installation contained 2,192 forensic facts. Each fact had a unique \texttt{FactID} and a unique source record identifier. Citation verification covered all 2,192 facts. The final experiment evidence register and the LLM supply register each contained 2,192 rows.

The evidence-supply corpus and the fact-recall universe had different roles. The models received the accepted evidence package derived from the 2,192 supply rows. A later evaluation stage defined 27 authorized report-level facts for fact-recall and omission analysis. These 27 facts formed an evaluation universe. They were not a training set and did not represent the complete model input.

\subsubsection{Physical Device Case}
\label{subsubsec:lg_construction}

The physical corpus was assigned the case identifier \texttt{LG\_PHONE\_001}. It was constructed from a verified acquisition of an LG LGLS751 device running Android 5.1.1. The acquisition identifier was

\begin{center}
\longid{LG\_PHONE\_001\_ACQ\_20260627T100134907854Z}.
\end{center}

The accepted acquisition contained 505 verified files with a total logical size of 971,085,117 bytes. The initial inventory recorded 276 files from internal shared storage, 217 files from the system partition, and 12 files from the \texttt{DATA\_ANR} area. These source-area counts account for all 505 verified files.

Artifact examination and human review were completed without returning to the physical device or modifying the accepted acquisition. The consolidated human-review workspace contained 497 rows. It preserved source paths, file characteristics, extraction records, metadata decisions, and exclusion decisions. A controlled metadata-only stage then completed the final evidence characterization and freeze.

The accepted mapping package resolved 13 evidence-package requirements. It contained 65 allowlist rows, 62 source or extracted evidence rows, 13 normalized evidence rows, 72 exclusion rows, and five lineage rows. The source or extracted rows and the normalized rows were overlapping classifications. They were therefore not added together to estimate the number of unique evidence items.

The final experiment evidence register and the LLM supply register each contained 65 rows. The accepted reportability policy authorized zero reportable forensic facts and required zero citations. This outcome represented a controlled zero-finding boundary rather than missing evidence or missing evaluation data. The physical case was included to assess whether the models would remain within this restrictive reporting boundary or generate unsupported findings from technical and contextual records. Accordingly, this case was designed as an intentional stress test condition for boundary compliance rather than as a representative model of a typical operational investigation.

\subsubsection{Construction Controls and Final Data Roles}
\label{subsubsec:construction_controls}

Table~\ref{tab:dataset_construction_summary} summarizes the accepted construction outputs. The two cases are reported separately because they used different evidence units and reportability policies.

\begin{table*}[t]
\centering
\caption{Construction summary for the two Android forensic case corpora.}
\label{tab:dataset_construction_summary}
\begin{tabular}{p{0.24\textwidth}p{0.32\textwidth}p{0.34\textwidth}}
\hline
\textbf{Construction element} &
\textbf{\texttt{A13\_PUBLIC\_001}} &
\textbf{\texttt{LG\_PHONE\_001}} \\
\hline

Source &
Public Android 13 archive &
Physical LG LGLS751 acquisition \\

Verified acquisition content &
One preserved archive &
505 files and 971,085,117 logical bytes \\

Extraction and inventory &
388 of 388 ALEAPP parsers, 51,849 output files, 266 TSV files, and 221 database files &
276 internal shared-storage files, 217 system-partition files, and 12 \texttt{DATA\_ANR} files \\

Selected or mapped evidence &
26 selected artifacts, comprising 24 Tier 1A and two Tier 1B artifacts &
65 allowlist rows mapped to 13 evidence-package requirements \\

Normalized content &
26 JSONL files and 2,192 validated records &
13 normalized evidence rows within the accepted mapping \\

Human review &
26 batches, 650 field decisions, 52 timestamp decisions, and 702 total decisions &
497 consolidated review rows \\

Exclusion control &
22 excluded or conditional entries, with traceback and anomaly records preserved &
72 exclusion rows \\

Final evidence authority &
2,192 unique forensic facts with complete citation coverage &
65-row accepted evidence package under a zero-reportable-fact authority \\

LLM supply &
2,192 accepted supply rows &
65 accepted supply rows \\

Fact-recall universe &
27 authorized report-level facts &
Not applicable because the accepted reportable-fact count was zero \\

\hline
\end{tabular}
\end{table*}

Construction completion was calculated only when the accepted expected count and verified count represented the same unit:

\begin{equation}
\mathrm{CompletionRate} =
\frac{N_{\mathrm{verified}}}
{N_{\mathrm{expected}}}
\times 100.
\label{eq:construction_completion}
\end{equation}

For \texttt{A13\_PUBLIC\_001}, parser completion was 388/388 (100\%), artifact normalization was 26/26 (100\%), human-review completion was 26/26 (100\%), and fact-citation coverage was 2,192/2,192 (100\%). For \texttt{LG\_PHONE\_001}, acquisition-file verification was 505/505 (100\%), evidence-requirement resolution was 13/13 (100\%), and experiment-supply registration was 65/65 (100\%). Citation coverage was not applicable to the physical case because the accepted citation denominator was zero.

The final model input consisted of two separate frozen case packages. The public package contained 2,192 accepted evidence-supply rows. The physical package contained 65 accepted evidence-supply rows under a zero-reportable-fact boundary. These values were not combined into a single dataset size because they represented different evidence units and case policies. After report generation, the six accepted reports were decomposed into 612 atomic claims. These claims formed the evaluation dataset used for reliability scoring.

\paragraph{Ethics and data handling}
The \texttt{LG\_PHONE\_001} device was owned by one of the
authors and was acquired with the owner's authorization for
this controlled study. No personally identifying device content
is reported or publicly released, and identifying information
associated with third parties was excluded from publication.
The public case used the openly available Digital Corpora
Android 13 dataset. The study did not recruit or interact with
external participants.

\section{Methodology}
\label{sec:methodology}

\subsection{Study Design}
\label{subsec:study_design}

This study used a controlled comparative design to evaluate the factual reliability of LLM-assisted Android forensic reports. The experiment included two forensic cases and three pretrained models: Gemini, Llama, and Qwen. Each model produced one accepted report for each case, resulting in six reports.

Reliability was evaluated at the atomic-claim level. Each report was decomposed into individual factual propositions and checked against the controlling case authority. Claim content, citation validity, and fact recall were evaluated separately. The analysis was descriptive and did not involve model training, parameter updates, statistical significance testing, or a general model ranking.

\subsection{Architecture and Evidence Control}
\label{subsec:architecture}

Figure~\ref{fig:methodology_architecture} summarizes the workflow. Each case began with a frozen evidence authority. A controlled evidence package and reporting prompt were supplied to each model. The generated reports were decomposed into atomic claims and evaluated against the accepted evidence, citation, and reportability authorities. All labels underwent independent rechecking, and disputed labels were resolved before creation of the final scoring dataset.

The verified evidence records were the primary authority. Human reference reports supported interpretation of the reporting boundary but were not supplied to the models and could not override the underlying evidence.

\begin{figure*}[t]
    \centering
    \safeincludegraphics[width=\textwidth]{figures/architecture.png}
    \caption{Controlled workflow for evidence-grounded LLM-assisted Android forensic reporting.}
    \label{fig:methodology_architecture}
\end{figure*}

\subsection{Forensic Evidence Cases}
\label{subsec:evidence_cases}

The experiment used two case-specific evidence packages.

\texttt{A13\_PUBLIC\_001} was derived from a public Android 13 dataset. Its accepted evidence package contained 2,192 supply rows. A separate set of 27 authorized report-level facts was used only for fact-recall and omission evaluation.

\texttt{LG\_PHONE\_001} was derived from a physical LG LGLS751 device running Android 5.1.1. Its accepted evidence package contained 65 supply rows. The controlling policy authorized zero reportable forensic facts and zero required citations.

The cases were not treated as interchangeable samples. Each retained its own evidence authority, reportability policy, reference report, and evaluation records.

\subsection{Evidence Preparation and Data Roles}
\label{subsec:model_data_roles}

Evidence preparation preceded all model activity. Original and derived records were preserved through case identifiers, source paths, file sizes, SHA256 values, manifests, lineage registers, and read-only packages. Structured records retained the source artifact, record identifier, field value, timestamp meaning, and human-review decision.

The models were used only for inference. No model was trained or fine-tuned, and the experiment contained no training set, validation set, or supervised test split. Table~\ref{tab:data_roles} defines the role of each data group.

\begin{table*}[t]
\centering
\caption{Data roles in the experiment.}
\label{tab:data_roles}
\begin{tabular}{p{0.18\textwidth}p{0.72\textwidth}}
\hline
\textbf{Data role} & \textbf{Use in the study} \\
\hline
Training data &
None. No model parameters were updated. \\

Validation data &
None. No model selection or hyperparameter tuning was performed. \\

Model input &
The frozen case evidence package and case-specific reporting prompt. The public case supplied 2,192 evidence rows. The physical case supplied 65 evidence rows under a zero-reportable-fact boundary. \\

Evaluation authority &
Frozen evidence records, reportability decisions, citation authorities, case policies, and human reference reports. These materials were used after report generation. \\

Evaluation data &
Six accepted reports containing 612 atomic claims: 547 from the public case and 65 from the physical case. \\
\hline
\end{tabular}
\end{table*}

Within each case, all three models received the same evidence boundary. They did not receive another model's report, the human reference report, final evaluation labels, or evidence from the other case.

\subsection{Human Reference Reports}
\label{subsec:human_reference}

A human reference report was prepared and independently verified for each case before claim evaluation. The public-case report recorded the authorized findings, evidence references, and limitations. The physical-case report recorded the accepted zero-reportable-fact state. Final claim decisions remained tied to the primary evidence authority.

\subsection{Report Generation}
\label{subsec:report_generation}

The accepted model identities and case-specific execution
settings are reported in Table~\ref{tab:model_configuration}.
The complete frozen prompt and exact request metadata were
preserved in the accepted experiment packages.

\begin{table*}[t]
\centering
\caption{Accepted model and generation configurations.}
\label{tab:model_configuration}
\scriptsize
\setlength{\tabcolsep}{4pt}
\renewcommand{\arraystretch}{1.06}
\begin{tabular}{p{0.06\textwidth}
                p{0.29\textwidth}
                p{0.23\textwidth}
                p{0.12\textwidth}
                p{0.06\textwidth}
                p{0.13\textwidth}}
\toprule
\textbf{Model} &
\textbf{Exact identifier} &
\textbf{Provider: LG / A13} &
\textbf{Date: LG / A13} &
\textbf{Temp.} &
\textbf{Maximum output tokens: LG / A13} \\
\midrule

Gemini &
\texttt{gemini-3.1-flash-lite} &
Google Gemini API / Google Gemini API &
15 Jul. / 16 Jul. 2026 &
0.0 &
65,536 / 65,536 \\

Llama &
\texttt{meta-llama/Llama-3.1-8B-Instruct} &
HF--Nscale / HF--Nscale &
16 Jul. / 17 Jul. 2026 &
0.0 &
4,096 / 14,848 \\

Qwen &
\texttt{Qwen/Qwen3-235B-A22B-Instruct-2507} &
HF--DeepInfra / HF--Novita &
16 Jul. / 17 Jul. 2026 &
0.0 &
14,848 / 14,848 \\

\midrule
\multicolumn{6}{p{1.00\textwidth}}{
\textit{Shared controls:} frozen reporting prompt v1.0;
top-$p$ and top-$k$ were not explicitly supplied; one text
response was requested; tools and external grounding were
disabled; no seed field was supplied in the accepted requests.
HF denotes Hugging Face Inference Providers.} \\
\bottomrule
\end{tabular}
\end{table*}

The frozen prompt required each model to use only the supplied evidence, provide evidence references, avoid unsupported interpretation, and state limitations when the evidence was insufficient. Failed attempts remained in the project history and were not substituted for accepted reports.

\subsection{Atomic Claim Extraction}
\label{subsec:claim_extraction}

Each accepted report was decomposed into atomic claims. An atomic claim represented one evaluable factual proposition. Compound statements were split only when the resulting claims preserved the original meaning and necessary context. Headings and other non-evidentiary text were recorded separately.

The extraction process preserved claim identifiers, exact text spans, report references, model identifiers, and lineage records. A separate verification stage checked claim boundaries, reconstruction, exclusions, and required context.

The accepted corpus contained 612 claims. Table~\ref{tab:claim_corpus} shows their distribution.

\begin{table}[t]
\centering
\caption{Accepted atomic claim corpus.}
\label{tab:claim_corpus}
\begin{tabular}{lrrrr}
\hline
\textbf{Case} & \textbf{Gemini} & \textbf{Llama} & \textbf{Qwen} & \textbf{Total} \\
\hline
\texttt{A13\_PUBLIC\_001} & 203 & 142 & 202 & 547 \\
\texttt{LG\_PHONE\_001} & 20 & 20 & 25 & 65 \\
\hline
Total & 223 & 162 & 227 & 612 \\
\hline
\end{tabular}
\end{table}

\subsection{Claim Labeling}
\label{subsec:claim_labeling}

Each claim was evaluated against the controlling case authority using the labels in Table~\ref{tab:label_policy}. The decision precedence was \(X > C > O > S > U\).

\begin{table}[t]
\centering
\caption{Claim labels used in final scoring.}
\label{tab:label_policy}

\small
\setlength{\tabcolsep}{2pt}
\renewcommand{\arraystretch}{1.10}

\begin{tabular}{
@{}
p{0.09\columnwidth}
p{0.27\columnwidth}
p{0.60\columnwidth}
@{}
}
\toprule
\textbf{Code} & \textbf{Name} & \textbf{Decision basis} \\
\midrule

S & Supported &
The authority supports the material factual content. \\

U & Unsupported &
The authority provides insufficient support and contains no controlling contradiction. \\

C & Contradicted &
The authority contains an incompatible value, state, scope, or negation. \\

O & Overstated or incomplete &
The claim exceeds the authority or omits a required qualifier. \\

X & Excluded &
The text is non-substantive and outside factual evaluation. \\

\bottomrule
\end{tabular}

\end{table}

Citation validity and fact recall were evaluated separately from claim content. A missing authorized fact was recorded as an omission rather than assigned to an existing claim. For example, if the evidence authority directly supports an artifact value, a claim reporting that value is labeled Supported. A claim that adds an interpretation not established by the authority is labeled Unsupported, while a claim that conflicts with the recorded value is labeled Contradicted. A claim that is partly supported but extends beyond the accepted evidence boundary is labeled Overstated or incomplete.

\subsection{Reliability Control and Dispute Resolution}
\label{subsec:adjudication}

The evaluation used three controls: restricted evidence input, atomic claim verification, and independent label rechecking. The original and independent assessments agreed on 494 of 612 claims. The remaining 118 claims entered evidence-based dispute resolution, during which the competing labels and rationales were concealed until a decision was recorded.

Human evaluation was conducted by three doctoral researchers, all authors of this study, with training in digital forensics and cybersecurity. They collectively performed the initial labeling, independent rechecking, and dispute resolution. The final scoring dataset retained 494 agreement labels and replaced 118 disputed labels with their accepted resolutions. Claim text, identifiers, evidence boundaries, and case policies remained unchanged.

\subsection{Evaluation Measures}
\label{subsec:evaluation_measures}

Let \(S\), \(U\), \(C\), \(O\), and \(X\) denote supported, unsupported, contradicted, overstated or incomplete, and excluded claims. Excluded claims were removed from substantive denominators:

\begin{equation}
N_{\mathrm{sub}} = S + U + C + O.
\label{eq:substantive_count}
\end{equation}

The claim-content measures were

\begin{equation}
\mathrm{GroundingRate} =
\frac{S}{N_{\mathrm{sub}}},
\label{eq:grounding_rate}
\end{equation}

\begin{equation}
\mathrm{UnsupportedRate} =
\frac{U}{N_{\mathrm{sub}}},
\label{eq:unsupported_rate}
\end{equation}

\begin{equation}
\mathrm{ContradictionRate} =
\frac{C}{N_{\mathrm{sub}}},
\label{eq:contradiction_rate}
\end{equation}

\begin{equation}
\mathrm{HallucinationRate} =
\frac{U+C}{N_{\mathrm{sub}}},
\label{eq:hallucination_rate}
\end{equation}

and

\begin{equation}
\mathrm{OverclaimRate} =
\frac{O}{N_{\mathrm{sub}}}.
\label{eq:overclaim_rate}
\end{equation}

Citation validity was calculated as

\begin{equation}
\mathrm{CitationValidityRate} =
\frac{N_{\mathrm{valid\ citations}}}
     {N_{\mathrm{citation\ required}}},
\label{eq:citation_validity}
\end{equation}

and fact recall was calculated as

\begin{equation}
\mathrm{FactRecall} =
\frac{N_{\mathrm{recovered\ authorized\ facts}}}
     {N_{\mathrm{authorized\ facts}}}.
\label{eq:fact_recall}
\end{equation}

A measure was reported as not applicable when its accepted denominator was zero. Report-level rates were calculated from report counts. Model-level and case-level rates were calculated from pooled counts rather than averaged report percentages.

\subsection{Verification and Reproducibility}
\label{subsec:verification}

The experiment used deterministic, phase-controlled Python 3.13.14 workflows. Each action preserved its execution script, source-authority snapshot, validation records, output manifest, package aggregate, and SHA256 values. Completed packages were frozen as read-only.

Each accepted stage required technical execution, an independent read-only verification pass, analyst review, and explicit acceptance. The final scoring package preserved one lineage row per claim and linked every final label to either an agreement decision or an accepted dispute resolution.

\section{Results}
\label{sec:results}

\subsection{Final Scoring Corpus}
\label{subsec:final_scoring_corpus}

The final scoring corpus contained 612 atomic claims from six accepted reports. The original and independent assessments agreed on 494 claims. The remaining 118 claims received the accepted labels produced through controlled dispute resolution. The final corpus contained 424 supported claims, 163 unsupported claims, four contradicted claims, four overstated or incomplete claims, and 17 excluded claims.

Table~\ref{tab:final_labels} reports the final claim-label counts by case and model, together with the pooled model totals.

\begin{table*}[!t]
\centering
\caption{Final claim-label counts by case and model, including pooled model totals.}
\label{tab:final_labels}
\small
\setlength{\tabcolsep}{5pt}
\renewcommand{\arraystretch}{1.00}

\begin{tabular*}{\textwidth}
{@{\extracolsep{\fill}}llrrrrrr@{}}
\toprule
\textbf{Case} & \textbf{Model} &
\textbf{S} & \textbf{U} & \textbf{C} &
\textbf{O} & \textbf{X} & \textbf{Total} \\
\midrule

\multirow{3}{*}{\texttt{A13\_PUBLIC\_001}}
& Gemini & 156 & 43 & 3 & 1 & 0 & 203 \\
& Llama  & 85  & 56 & 0 & 1 & 0 & 142 \\
& Qwen   & 175 & 26 & 0 & 1 & 0 & 202 \\
\addlinespace[2pt]

\multirow{3}{*}{\texttt{LG\_PHONE\_001}}
& Gemini & 4 & 15 & 0 & 1 & 0 & 20 \\
& Llama  & 0 & 3  & 0 & 0 & 17 & 20 \\
& Qwen   & 4 & 20 & 1 & 0 & 0 & 25 \\
\midrule

\multirow{3}{*}{\textbf{Pooled}}
& Gemini & 160 & 58 & 3 & 2 & 0 & 223 \\
& Llama  & 85  & 59 & 0 & 1 & 17 & 162 \\
& Qwen   & 179 & 46 & 1 & 1 & 0 & 227 \\
\bottomrule
\end{tabular*}
\end{table*}

\subsection{Model-Level Outcomes}
\label{subsec:model_level_outcomes}

\begin{figure}[t]
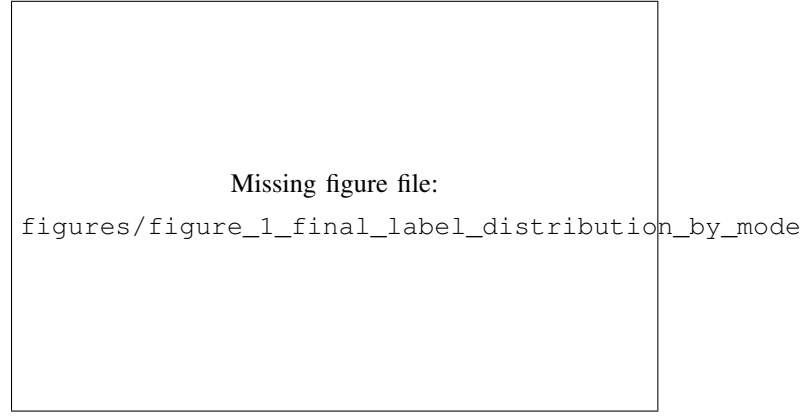

    \centering
    \safeincludegraphics[width=\linewidth]
    {figures/figure_1_final_label_distribution_by_model.png}
    \caption{Final claim-label counts pooled across both cases for each model.}
    \label{fig:final_labels_model}
\end{figure}

Figure~\ref{fig:final_labels_model} and the pooled rows of Table~\ref{tab:final_labels}  summarize the final labels across both cases. Gemini generated 223 claims, Llama generated 162 claims, and Qwen generated 227 claims. Their supported-claim counts were 160, 85, and 179, respectively. Unsupported-claim counts were 58 for Gemini, 59 for Llama, and 46 for Qwen. Contradicted and overstated or incomplete claims occurred less frequently but remained present in the final outputs.

These results describe the six accepted reports and do not establish a general ranking among the models. The models produced different numbers of claims, and the two cases imposed different evidence and reportability conditions.

\subsection{Case-Level Outcomes}
\label{subsec:case_level_outcomes}

\begin{figure}[t]
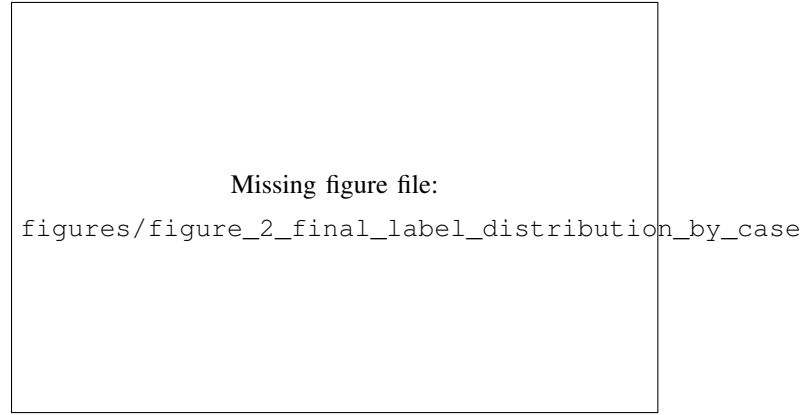

    \centering
    \safeincludegraphics[width=\linewidth]
    {figures/figure_2_final_label_distribution_by_case.png}
    \caption{Final claim-label counts pooled across the three
    models within each case.}
    \label{fig:final_labels_case}
\end{figure}

Figure~\ref{fig:final_labels_case} summarizes the final claim-label distribution by case. The public Android case contained 547 substantive claims: 416 supported, 125 unsupported, three contradicted, and three overstated or incomplete. The physical-device case contained 48 substantive claims: eight supported, 38 unsupported, one contradicted, and one overstated or incomplete; 17 additional claims were excluded. Using \(N_{\mathrm{sub}} = S + U + C + O\), the public case achieved a grounding rate of 76.05\% and an observed hallucination rate of 23.40\%, while the physical-device case achieved 16.67\% and 81.25\%, respectively. Because the cases used distinct reporting authorities, these rates should not be interpreted as a direct comparison between public and physical evidence.

Across models, grounding rates were 71.75\% for Gemini,
58.62\% for Llama, and 78.85\% for Qwen. Their corresponding
hallucination rates were 27.35\%, 40.69\%, and 20.70\%.
Across all 595 substantive claims, the overall grounding and
hallucination rates were 71.26\% and 28.07\%, respectively.

\subsection{Citation Validity and Fact Recall}
\label{subsec:citation_recall_results}

Table~\ref{tab:citation_recall} reports citation validity and
fact recall for the public Android case. Gemini achieved
88.89\% citation validity with 168 valid citations from 189
required citations. Llama achieved 0\% with no valid citations
from 135 required citations. Qwen achieved 76.72\% with
145 valid citations from 189 required citations.

All three models recovered all 27 authorized report-level
facts. Complete fact recall therefore did not imply valid
evidence attribution. Citation validity and fact recall were
not applicable to the physical-device case because its
accepted authority contained zero reportable facts and zero
required citations.

\begin{table*}[!t]
\centering
\caption{Citation validity and fact recall by report.}
\label{tab:citation_recall}
\small
\setlength{\tabcolsep}{6pt}
\renewcommand{\arraystretch}{1.00}

\begin{tabular*}{\textwidth}
{@{\extracolsep{\fill}}llccccc@{}}
\toprule
\textbf{Case} &
\textbf{Model} &
\textbf{Valid} &
\textbf{Required} &
\textbf{Validity (\%)} &
\textbf{Recovered} &
\textbf{Authorized} \\
\midrule

\multirow{3}{*}{\texttt{A13\_PUBLIC\_001}}
& Gemini & 168 & 189 & 88.89 & 27 & 27 \\
& Llama  & 0   & 135 & 0.00  & 27 & 27 \\
& Qwen   & 145 & 189 & 76.72 & 27 & 27 \\

\addlinespace[2pt]

\multirow{3}{*}{\texttt{LG\_PHONE\_001}}
& Gemini & N/A & N/A & N/A & N/A & N/A \\
& Llama  & N/A & N/A & N/A & N/A & N/A \\
& Qwen   & N/A & N/A & N/A & N/A & N/A \\

\bottomrule
\end{tabular*}
\end{table*}

\subsection{Result Boundaries}
\label{subsec:result_boundaries}

The results are limited to six fixed reports generated from two controlled Android forensic cases. They do not establish causal effects, overall model superiority, output stability across repeated generations, or generalization beyond the evaluated evidence packages, models, prompts, and scoring controls.

\section{Discussion}
\label{sec:discussion}

\subsection{Evidence Grounding and Observed Hallucination}
\label{subsec:discussion_rq1}

Evidence grounding did not prevent unsupported model output. Of the 612 accepted claims, 424 were supported, 163 were unsupported, four were contradicted, four were overstated or incomplete, and 17 were excluded. These results show that a controlled evidence package alone cannot guarantee evidentiary compliance.

The physical-device case demonstrated this limitation most clearly. Although its authority contained zero reportable forensic facts, the models produced 65 claims. Only eight were supported, while 38 were unsupported and one was contradicted. The framework therefore does not eliminate hallucination. It detects claims that exceed the accepted evidence boundary before they enter a final report. This finding is consistent with earlier work on errors in LLM-assisted forensic reporting \citep{michelet2024chatgpt}.

\subsection{Differences Among the Models}
\label{subsec:discussion_rq2}

The three models produced different claim distributions. Qwen generated 179 supported and 46 unsupported claims. Gemini generated 160 supported and 58 unsupported claims. Llama generated 85 supported and 59 unsupported claims, with 17 additional claims excluded from substantive evaluation.

These counts describe the six accepted reports and do not establish a general model ranking. The models produced different report lengths and operated under two different case policies. Their outputs must therefore be evaluated at the report and claim levels rather than through a single overall score. This interpretation is consistent with task-specific forensic evaluation against accepted evidence \citep{Studiawan2025}.

\subsection{Fact Recall and Citation Reliability}
\label{subsec:discussion_rq3}

All three models recovered the 27 authorized report-level facts in the public case. Citation validity, however, differed substantially. Gemini produced 168 valid citations from 189 required citations. Qwen produced 145 valid citations from 189 required citations. Llama produced no valid citations from 135 required citations.

Complete fact recall therefore did not imply reliable evidence attribution. A report may include every authorized fact while still containing invalid or insufficient citations. Fact recall, claim support, and citation validity should consequently remain separate evaluation dimensions. Citation failures can also arise for different reasons. For example, a generated reference may point to an existing evidence item using an incorrect identifier or citation form, whereas another reference may identify evidence that is not present within the accepted authority. These failure modes require separate investigator verification rather than treating all invalid citations as equivalent. This result supports prior findings that improved source attribution does not remove the need for citation verification \citep{Sharma2025}.

\subsection{Differences Between the Two Cases}
\label{subsec:discussion_rq4}

The public case contained 416 supported claims among 547 claims. The physical-device case contained eight supported claims among 65 claims. This difference should not be interpreted as a direct comparison between public and physical Android evidence.

The cases used different reporting authorities. The public case had 27 authorized report-level facts within a larger evidence package. The physical case imposed a zero-reportable-fact boundary. Its high unsupported count shows that the models often converted technical or contextual records into claims that exceeded the accepted authority. The presence of files, metadata, timestamps, or identifiers does not by itself establish user identity, intent, ownership, or physical action.

\subsection{Implications for Forensic Reporting}
\label{subsec:practical_implications}

The results support using LLMs as drafting assistants rather than autonomous report generators. Investigators must retain control over evidence boundaries, claim interpretation, citation validation, and final approval. This requirement is consistent with prior work emphasizing transparency and human validation in forensic AI \citep{Vasilaras2024,Khalid2024}.

The proposed workflow combines restricted evidence input, atomic claim extraction, claim-to-evidence verification, independent rechecking, and dispute resolution. These controls do not guarantee error-free reports, but they make unsupported, contradicted, and overstated claims identifiable and traceable before acceptance. The framework is therefore intended as an evaluation and verification mechanism rather than an automated method for preventing hallucinations or citation errors during generation.

\subsection{Interpretive Boundary}
\label{subsec:discussion_boundary}

The findings are limited to six fixed reports generated from two controlled Android forensic cases. They do not establish overall model superiority, performance across other devices or artifacts, stability across repeated generations, reduced investigator workload, improved legal outcomes, or elimination of hallucination.

Within this scope, the results support a narrower conclusion: pretrained LLMs can recover authorized forensic facts, but they can also generate unsupported claims and unreliable citations. Defensible use therefore requires frozen evidence authorities, claim-level verification, preserved lineage, independent review, and final investigator approval.

\section{Limitations}
\label{sec:limitations}

This study has four main limitations. First, it evaluates only two Android cases, so the findings may not generalize across devices, operating-system versions, applications, acquisition methods, or investigative settings. Second, the cases use different reporting boundaries: the public case is limited to a bounded Google Voice scope, while the physical-device case is an intentional zero-reportable-fact stress test. These conditions support controlled evaluation but limit direct case-to-case and broader operational comparisons. Third, each model produced one accepted report per case. Repeated-generation variance, provider changes, and model updates were not assessed; therefore, the results do not establish statistical significance or a general model ranking. Finally, claim extraction, labeling, rechecking, and dispute resolution require human judgment. Although preserved policies and lineage improve consistency and traceability, reviewer effort and operational scalability were not measured. The study also does not assess usability or legal admissibility.

\section{Future Work}
\label{sec:future_work}

Future work should evaluate the framework on a broader set of Android devices, operating-system versions, acquisition methods, applications, and artifact types. Repeated report-generation runs are also needed to measure output stability and support statistical comparison across models.

The same evidence authority and claim-level evaluation procedure could be used to compare prompting strategies, retrieval-assisted reporting, and domain-specific fine-tuning. Automated claim extraction and preliminary evidence matching may reduce review effort, but unsupported, contradicted, and overstated claims should remain subject to investigator confirmation.

Further studies should examine citation validity, completeness, and evidence proximity as separate measures. Practitioner studies could also assess correction effort, usability, and reviewer confidence. Extensions to iOS and other forensic domains should retain verified evidence boundaries, preserved lineage, independent review, and final human approval.

\section{Conclusion}
\label{sec:conclusion}

This study evaluated six LLM-generated Android forensic reports produced by Gemini, Llama, and Qwen from two controlled evidence cases. The reports were decomposed into 612 atomic claims and assessed against frozen, human-reviewed evidence authorities.

The final corpus contained 424 supported, 163 unsupported, four contradicted, four overstated or incomplete, and 17 excluded claims. All three models recovered the 27 authorized report-level facts in the public case, but citation validity differed substantially. The physical-device case further showed that a zero-reportable-fact boundary did not prevent the generation of unsupported findings.

These results demonstrate that evidence-constrained prompting alone is insufficient for reliable forensic reporting. LLMs may assist with report drafting, but their outputs require claim-level verification, citation checking, preserved evidence lineage, independent review, and final investigator approval. The proposed framework provides a controlled method for identifying statements that are supported, require correction, or must be excluded from the final report.

\section*{AI Disclosure}

OpenAI ChatGPT was used to assist with restructuring and
language revision of the abstract and Sections I-IX.
Grammarly was used for grammar and clarity editing. Neither
tool conducted the forensic examination, generated or modified
experimental data, executed the evaluation, assigned final
labels, or computed the accepted results. The authors verified
the correctness, originality, numerical values, and references
of all content.

\section*{Data and Code Availability}

The public Android 13 dataset is available through Digital
Corpora. The frozen prompts, evaluation definitions, aggregate
results, and metric-generation scripts will be released in an
anonymized artifact repository after acceptance. The physical
device evidence, private records, credentials, and identifying
metadata will not be released.

\bibliographystyle{IEEEtranN}
\bibliography{references}

\end{document}